\documentclass[12pt]{article}
\usepackage[utf8]{inputenc}
\usepackage[left=1in,right=1in,top=1in,bottom=1in]{geometry}
\usepackage{booktabs} 
\usepackage{authblk} 
\usepackage{graphicx} 
\usepackage{comment}
\usepackage{breakcites}
\usepackage[bookmarks=false]{hyperref}
\usepackage{amsmath}
\usepackage{amssymb}
\usepackage[dvipsnames]{xcolor}
\usepackage{lineno}
\usepackage{tabularx}

\title{Strength of minority ties: the role of homophily and group composition in a weighted social network}

\author[1]{J. R. Nicolás-Carlock\footnote{Correspondence: jnicolas@unam.mx}}
\author[1]{D. Boyer}
\author[2]{S. E. Smith-Aguilar}
\author[2]{G. Ramos-Fernández}

\affil[1]{\normalsize Instituto de Física, Universidad Nacional Autónoma de México, México}
\affil[2]{Instituto de Investigaciones en Matemáticas Aplicadas y en Sistemas, Universidad Nacional Autónoma de México, México}

\date{}
\begin{document}

\maketitle

\begin{abstract}

Homophily describes a fundamental tie-formation mechanism in social networks in which connections between similar nodes occur at a higher rate than among dissimilar ones. In this article, we present an extension of the Weighted Social Network (WSN) model that, under an explicit homophily principle, quantifies the emergence of attribute-dependent properties of a social system. To test our model, we make use of empirical association data of a group of free-ranging spider monkeys in Yucatan, Mexico. Our homophilic WSN model reproduces many of the properties of the empirical association network with statistical significance, specifically, the average weight of sex-dependent interactions (female-female, female-male, male-male), the weight distribution function, as well as many weighted macro properties (node strength, weighted clustering, and weighted number of modules), even for different age group combinations (adults, subadults, and juveniles). Furthermore, by performing simulations with fitted parameters, we show that one of the main features of a spider monkey social system, namely, stronger male-male interactions over female-female or female-male ones, can be accounted for by an asymmetry in the node-type composition of a bipartisan network, independently of group size. The reinforcement of connections among members of minority groups could be a general structuring mechanism in homophilic social networks. 

\end{abstract}

\section*{Introduction}

Network-based methodologies offer a robust framework for the modeling and rigorous quantitative analysis of social structures \cite{wasserman1994social, smith2019, brask2021animal}. By their nature, social structures are complex, always evolving in response to endogenous factors and external conditions \cite{ramos2009association, ramos2018quantifying}. Consequently, the inferences that can be drawn using network approaches are as good as the quality of the available data, underlying network models, and sociological theories \cite{sosa2020network}. Despite these challenges, simple models of network formation have been proposed to understand the microscopic mechanisms that drive the evolution of social networks \cite{watts1998collective, barabasi1999emergence, bianconi2001competition, marsili2004rise, kossinets2006empirical}. Among these, the Weighted Social Network (WSN) model \cite{kumpula2007} stands out as a simple model that captures some of the features of real social systems, namely, weighted connections and the emergence of Granovetterian structures in which strongly connected communities connect among each other by weak links, the so-called \textsl{strength of weak ties} \cite{granovetter1973strength}.

The WSN model relies on two main mechanisms of tie-formation known as triadic closure or the formation of connections among shared neighbors (friends of friends), and focal closure or the formation of random connections. Nonetheless, associations and the formation of relationships in social systems also depend on explicit factors related to the nature and shared characteristics among individuals \cite{axelrod1997dissemination}. \textsl{Homophily}, in particular, is a sociological principle that establishes that individuals tend to associate and form relationships with similar others more than with dissimilar ones \cite{lazarsfeld1954friendship, mcpherson2001birds}. In social networks, this similarity can be understood in terms of node attributes, such as sex, age, ethnicity, religion, political views, education, occupation or popularity. Although there are positive consequences of homophily (e.g. tolerance, cooperation and innovation), there are also negative ones (e.g. discrimination, segregation and polarization). Therefore, understanding the role of homophily on the emergence of the dynamical and structural properties of social networks is of great relevance for societies \cite{sert2020segregation, karimi2018homophily, evtushenko2021paradox, khanam2023homophily}. 

In this article, we present an agent-based model for the weighted structure of social networks that considers controlled attribute-dependent interactions in order to understand the role of homophily on the emergent structural properties of social systems. To that end, we introduce an extension of the WSN model in which the classical tie-formation mechanisms are now governed by an explicit homophily principle based on node attributes. To test our model, we apply it to data from field observations of associations among a group of free-ranging spider monkeys living in a protected area in Yucatan, Mexico. The choice of this social system relies on the strong degree of sex-related homophily that it exhibits, with closer male-male relationships over female-male or female-female ones. Standard socio-ecological theory explains those patterns as a result of the different evolved strategies of the sexes: males cooperate to defend the range of a group of females, who in turn isolate themselves in order to avoid competition \cite{sterck1997evolution, aureli2008social, ramos2009association}. 

Even more, spider monkeys live in communities with a high degree of fission–fusion dynamics, where members frequently split and merge into subgroups, adjusting to the availability and distribution of resources (scarcity leading to small subgroups, while abundance to larger subgroups), and predation pressure. The high degree of fission–fusion dynamics and migration events influence the chance of interaction among community members,which in turn impacts the number and quality of the social relationships. Each partner in a relationship is also involved in relationships with other group members, so that each relationship is part of a network of relationships or social structure conforming sex-segregated groups, with females being the ``less social'' sex. This segregation is mediated by age, as  young males tend to associate more strongly with their mothers and other females but when approaching sexual maturity they progressively associate more exclusively with adult males \cite{ramos2009association}. As such, our network model aims to explore the role of node attributes and relative sex composition on the emergence of the social structure of spider monkeys, specifically, sex-dependent interactions. To further explore more general scenarios, we performed numerical simulations with fitted parameters considering different network sizes and ``sex'' compositions modelled as bipartisan networks with controlled majority and minority groups.

The article is structured as follows. First, we introduce the extended WSN model along with the network metrics and statistical fitting method to be employed. Second, we describe the nature of the data and the corresponding empirical association network. Subsequently, the results are presented in two parts: the first relates to the application of the statistical fitting method for different age group combinations; in the second part, we explore numerically the predictions of the model according to the fitted parameters for different node-type compositions and network sizes. We conclude with a discussion of the results.

\section*{Methods}

\subsection*{Extended WSN model}

\begin{figure}[t!]
\includegraphics[width=1.0\textwidth]{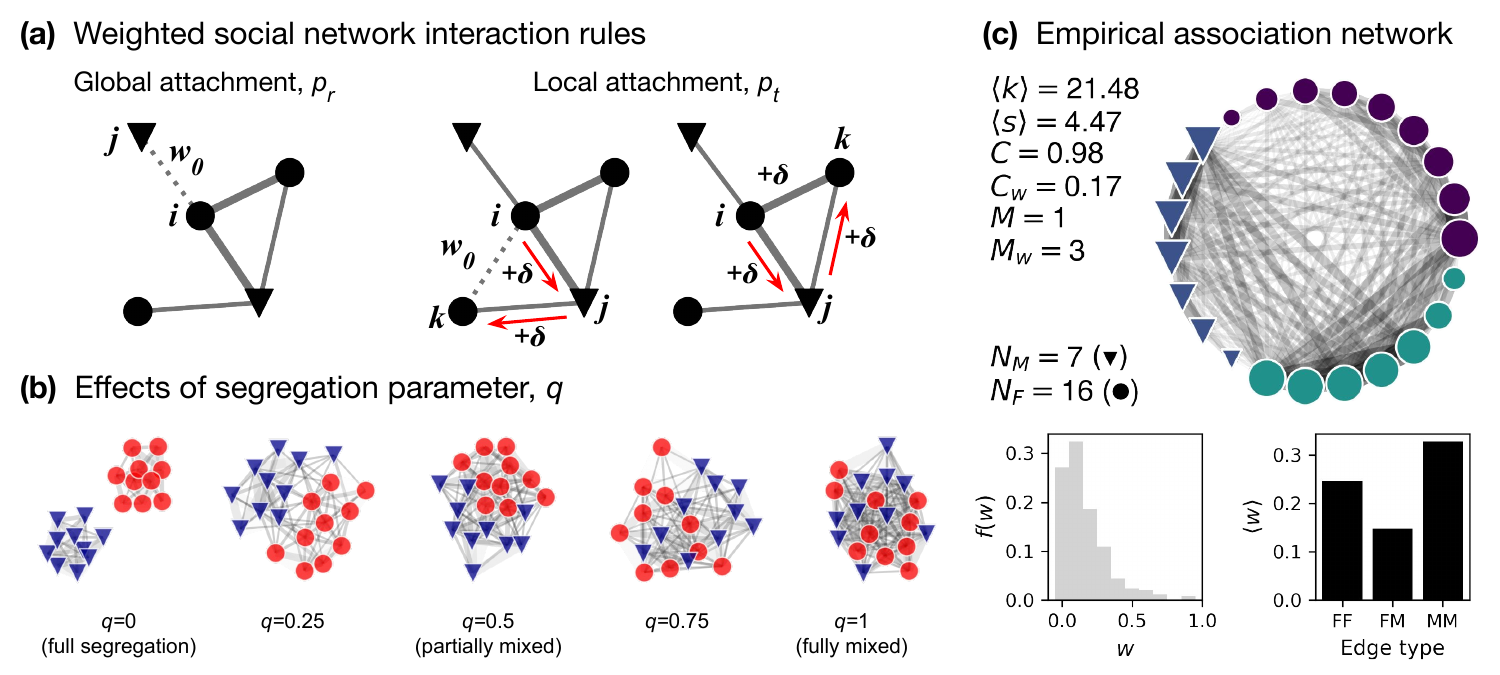}
\caption{\small \textbf{Extended WSN model and empirical network.} (a) Schematics of the WSN model interaction rules: global and local attachment (no deletion is considered). (b) Qualitative effect of the segregation parameter $q$ over the structure of the networks. (c) Visualization of the empirical association network (males/females as triangles/circles, respectively). Node size is proportional to strength, edge width to weight, and node colors indicate community membership (\textit{via} the Luvain method). Also indicated: average degree, $\langle k \rangle$, and strength, $\langle s \rangle$; unweighted and weighted average clustering coefficient, $C$ and $C_w$, respectively; unweighted and weighted number of communities, $M$ and $M_w$, respectively; weight distribution function, $f(w)$, and average weight per edge-type, $\langle w \rangle$.}
\label{fig:fig1}
\end{figure}

The original WSN model \cite{kumpula2007} has been implemented considering different variations, for example, link deletion and aging mechanisms \cite{murase2015}, multilayer features \cite{murase2014}, and extreme homophily \cite{murase2019}. Here, we introduce an extension that considers nodes with different attributes and continuous attribute-dependent interactions. Based on social network theory \cite{kossinets2006empirical}, the network evolution is governed by the following rules (see Fig. \ref{fig:fig1}a):

\begin{itemize}

    \item The Global Attachment (GA) process is controlled by the parameter $p_r\in[0,1]$ which defines the probability of random or focal closure connections. For each node $i$, a node $j$ (not connected to $i$) is randomly selected, then an edge of weight $w_{ij}=w_{ji}=w_0$ is created with probability $p_r$ between the two. If the degree of node $i$ is zero, then, an edge of weight $w_{ij}=w_0$ is created with $p_r=1$.

    \item The Local Attachment (LA) process is controlled by the parameter $p_t\in[0,1]$ which defines the probability of local or triadic closure connections. For each node $i$ (with degree different from zero), a node $j$ is selected among its neighbors with probability proportional to the weight $w_{ij}$, then another node $k$ is selected among the neighbors of node $j$ with probability proportional to $w_{jk}$. Here, three scenarios are possible: (\textit{i}) if node $j$ has node $i$ as its only neighbor, the weight $w_{ij}$ is incremented by a quantity $\delta$; (\textit{ii}) if node $k$ is also neighbor of node $i$, then, the weights $w_{ij}$, $w_{jk}$, and $w_{ki}$, are incremented by $\delta$; (\textit{iii}) if node $k$ is not a neighbor of node $i$, then, an edge of weight  $w_{ki}=w_0$ is created with probability $p_t$ and the weights $w_{ij}$ and $w_{jk}$ are incremented by $\delta$. The parameter $\delta$ hence reinforces the already existing edges and produces heterogeneous weights over time.

    \item Node/edge deletion. The deletion process is controlled by the parameter $p_d\in[0,1]$ which defines the probability of edge or node deletion. In the case of nodes, a random node $i$ is removed and replaced with a new node (with no connections) with probability $p_d$. In the case of edge deletion, edges are deleted with probability $p_d$. In any case, with $p_d=0$, edges or nodes are permanent.

\end{itemize}

We extended the WSN model to account for the homophilic effects of the nodes' attributes over the GA and LA tie-formation processes. This was done by considering nodes with a ``sex'' attribute, $\sigma=\{F,M\}$. The GA and LA processes are now governed by an explicit homophily principle that is modulated by a segregation parameter $q$ that takes continuous values in $[0,1]$, establishing the probability of connections between nodes of different sex. Note that this is quite different from binary or extreme homophily in which nodes can only connect if attributes match exactly \cite{murase2019}.

The GA and LA processes are modified as follows. In the case of GA, the focal player or node $i$ randomly chooses a candidate node $j$ (not connected with $i$); then, $i$ and $j$ connect with probability $p_r$ if the same attributes are shared ($\sigma_i=\sigma_j$), otherwise ($\sigma_i \neq \sigma_j$), they connect with a smaller probability $p_r\times q$. As before, edges are created with $w_{ij}=w_0$. In the case of LA, the local search is performed as before if two nodes involved in a connection have the same attribute: a node $i$ chooses a node $j$ among its neighbors with probability proportional to $w_{ij}$, then a node $k$ is selected among the neighbors of node $j$ with probability proportional to $w_{jk}$. If a pair of nodes does not have the same attributes, the corresponding action is taken with probability $q$. Namely, edges between nodes of different attributes are reinforced (when selected) with probability $q$, and triangles are formed with probability $p_t\times q$. 

In this way, $q\to 1$ leads to full mixing (original model), while for $q\to 0$ to full segregation, with non-trivial segregation in-between (see Fig. \ref{fig:fig1}b), a feature that is not considered in previous models.  

\subsection*{Empirical association network}

The empirical association network was originally presented in Ref. \cite{smith2019}. The data was collected between January 2013 and September 2014 from a well habituated group of spider monkeys living in the \textsl{Otoch Ma’ax yetel Kooh} protected area in Yucatan, Mexico. The group consisted of 23 monkeys (excluding individuals younger than 5 by the end of the study period): 7 males (M) and 16 females (F); in terms of simplified age class: 13 adults (A; age greater than 8 years), 8 subadults (SA; age 5-8 years), and 2 juveniles (J; age 3-5 years) which became subadults in the course of the study period. 

Edges represent associations based on aggregated data from scan sampling of spider-monkey subgroups, performed every 20 min. Scan samples comprise records of subgroup composition collected by experienced observers following one subgroup at a time for 4-8 daily hours during 244 days, hence we do not have the full record of all associations. Given the high degree of fission-fusion dynamics of the species, group members are found organized into subgroups which change their size and composition in the course of hours (a subgroup was defined as the set of individuals within 30 meters of another, see \cite{ramos2005vocal} for development and \cite{aureli2012subgroup} for validation of this definition of subgroup). A pair of individuals was considered to be associated if they were recorded in the same subgroup during a sample. The scan samples hence represent approximations to the instantaneous association patterns between individuals, with a temporal resolution of 20 min. The association index that defines the edges' weights is an aggregated quantity over the whole observation period and considers the proportion of scans two individuals were seen in the same subgroup, and takes values from 0 to 1. For comparison purposes with our network model, the association index is re-scaled by its maximum value. 

In Fig. \ref{fig:fig1}c, we show a visualization of the empirical network along with some network metrics (see Appendix). The study of this type of association network in spider monkeys as well as in other animal species have provided evidence that processes akin to the LA and GA rules as used in the WSN model \cite{kumpula2007} apply to these species as well (e.g. \cite{ramos2009association, fu2012evolution, pinter2014dynamics, borgeaud2016intergroup}).

\subsection*{Statistical fitting}

The network data is generated with the WSN model considering the parameters' values, $p_r\in[0.1,1.0]$, $p_t\in[0.0,1.0]$, and $q\in[0.0,1.0]$, with steps of $0.1$ each. For each parameter set $(p_r, p_t, q)$, an ensemble of $S=100$ networks is created. Each network starts from an initial set of $N$ disconnected nodes, following the GA and LA rules modulated by $q$, for a total of $T=100$ iterations. This ensures the formation of a high number of connections and sufficient edge reinforcement as in the empirical network. The probability of edge/node deletion is set to zero ($p_d=0$) to account for no individual loses (deaths or disappearances), nor loses in the count of interactions, similarly to the way the empirical network was constructed. We also set $\delta=w_0=1$. 

We rely on seven weighted metrics to fit the model: the average strength, $\langle s \rangle$, the average weighted clustering coefficient, $C_w$, the weighted number of communities, $M_w$, the weight distribution function, $f(w)$, and the average weight per edge-type, $\langle w \rangle_{FF}$, $\langle w \rangle_{FM}$, $\langle w \rangle_{MM}$ (see Appendix for definitions). Unweighted metrics such as the average degree, $\langle k\rangle$, average clustering coefficient, $C$, and number of unweighted communities, $M$, are somehow simple as their empirical values are very close to $N-1$, $1$, and $1$, respectively, and easily reproduced by the model. Therefore, we focus on the weighted metrics only, which take non-trivial values about the weighted social interactions. 

The statistical fitting of the model is performed for the seven aforementioned metrics considering three tests. For each set of parameters, $(p_r, p_t, q)$, the empirical metrics must fall within a significance range in order to be accepted. Hence, the set of parameters passing all the tests yield networks whose weighted metrics are statistically consistent with the empirical data. The tests are the following:

\begin{itemize}

    \item[] \textsl{Test 1 (T1) for the weighted macro metrics.} For each set of parameters, $(p_r, p_t, q)$, the empirical values, $\langle s \rangle$, $C_w$, and $M_w$, must fall within the significance range $[\alpha, 1-\alpha]$ of the corresponding statistical distribution, that is, a two-tail test with significance level $\alpha=0.05$.

    \item[] \textsl{Test 2 (T2) for the weight distribution function.} For each set of parameters, $(p_r, p_t, q)$, we performed a Kolmogorov-Smirnov (KS) test for the simulated weight distributions, ruling out the cases that have a $p$-value lower than 0.1 \cite{clauset2009power}.
    
    \item[] \textsl{Test 3 (T3) for the average weight according to edge-type.} For each set of parameters, $(p_r, p_t, q)$, the empirical values, $\langle w \rangle_{FF}$, $\langle w \rangle_{FM}$, and $\langle w \rangle_{MM}$, must fall within the significance range $[\alpha, 1-\alpha]$ of the corresponding statistical distribution, that is, a two-tail test with significance level $\alpha=0.05$.
    
\end{itemize}

Notice that these tests progressively probe the simulated network structures, from general properties (weighted macro metrics), through meso characteristics (weighted distribution function), to more specific features of interest (average weights according to edge-type).

\section*{Results}

\begin{figure}[t!]
\includegraphics[width=1.0\textwidth]{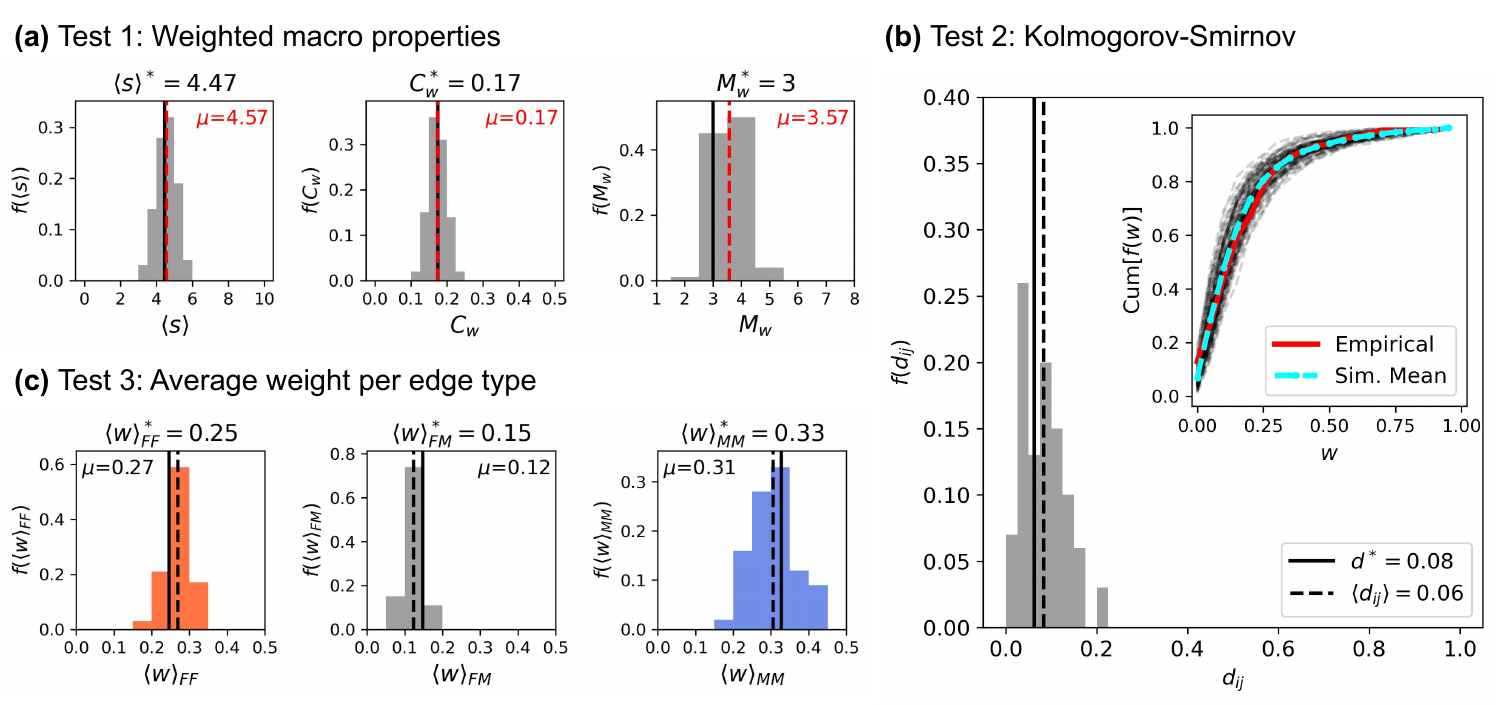}
\caption{\small \textbf{Statistical tests}. (a) Distribution functions of the strength, weighted clustering and weighted number of communities. The corresponding empirical values, $\langle s \rangle^*$, $C_w^*$, and $M_w^*$, are shown on top. The dotted red line indicates the mean of simulations, $\mu$, and the black line the corresponding empirical value. (b) Distribution of the Kolmogorov-Smirnov distances with the simulated mean, $d_{ij}$, and empirical mean, $d^*$, values indicated. In the inset, the corresponding cumulative distribution functions, Cum$[f(w)]$, with the mean and the empirical distribution functions indicated. (c) Distributions of the average weight per edge-type. The dotted line indicates the experimental mean, $\mu$, while the solid line the corresponding empirical value indicated with an asterisk (*) on top. The set of parameters used in these examples is $(p_r,p_t,q)=(0.7,0.5,0.7)$.}
\label{fig:fig2}
\end{figure}

\begin{figure}[t!]
\includegraphics[width=1.0\textwidth]{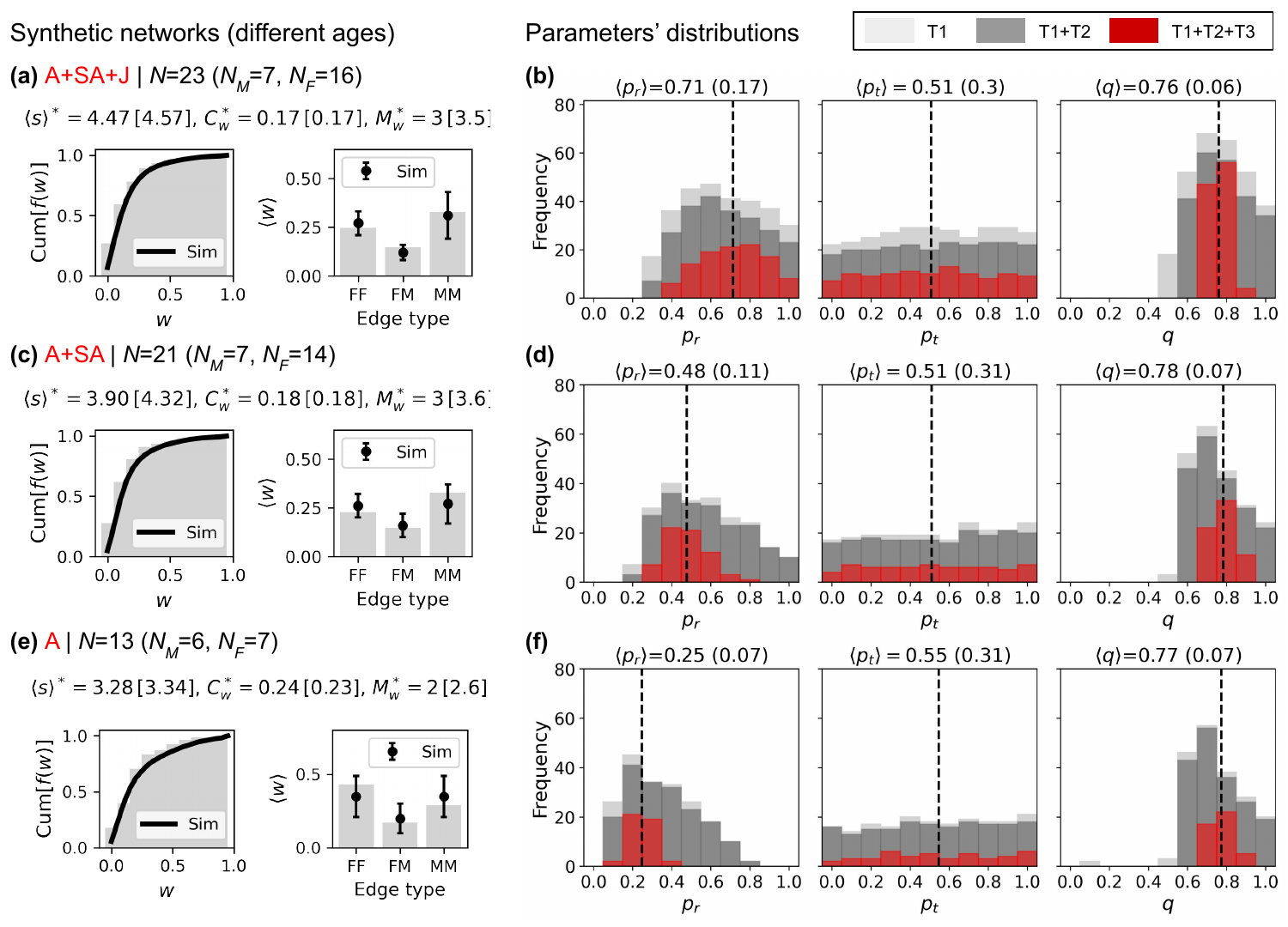}
\caption{\small \textbf{Distributions of parameters for different ages}. Results in (a)-(b) correspond to the association network with A (adults), SA (subadults), and J (juveniles). In (a), we show the empirical values ($\langle s \rangle^*$, $C_w^*$, $M_w^*$, Cum[$f(w)$], and $\langle w\rangle$ for each edge type) versus simulated results indicated in brackets. The simulated results were obtained using values close to averages, $\langle p_r \rangle$, $\langle p_t \rangle$, and $\langle q \rangle$, shown in (b) and indicated with the dotted line (the standard deviation is shown in the parenthesis). Results in (c)-(d), stand for A and SA, with $(p_r,p_t,q)=(0.5,0.5,0.8)$, and (e)-(f), only for A, with $(p_r,p_t,q)=(0.2,0.5,0.8)$.}
\label{fig:fig3}
\end{figure}

\subsection*{Statistical fitting to empirical association network}

We consider the following features of the empirical network for the simulations and statistical fitting: the number of nodes is set to $N=23$, with $N_M=7$ (males) and $N_F=16$ (females). Notice that this is not a symmetric composition of the number of males (a minority) relative to the females (the majority group). Also, the nodes include adults (A), subadults (SA), and juveniles (J). In general, the extended WSN model produces non-trivial numerical results for the weighted metrics, for example, in Fig. S1 (Supplementary Materials), we show the values for the weighted metrics, $\langle s\rangle$, $C_w$ and $M_w$, obtained from the model for different projections in parameter space. However, only those sets, $(p_r,p_t,q)$, that satisfy the significance tests are considered. In Fig. \ref{fig:fig2}, we present the results of a successful test considering the fitted parameters $(p_r,p_t,q)=(0.7,0.5,0.7)$.

In Fig. \ref{fig:fig2}a, the probability distributions of the weighted macro metrics recover the corresponding empirical observations with statistical significance. Recall that the tests are not independent among each other, therefore, these results show that the model is able to reproduce the non-trivial dependency among the average strength of connections, average weighted clustering, and the number of weighted modules, under the male/female composition asymmetry of the empirical network. In addition to recovering the average metrics, after the second test based on Kolmogorov-Smirnov distances, the model also recovers the empirical weight distribution (see Fig. \ref{fig:fig2}b) which encodes a more detailed structural characteristic of the empirical social structure, namely, the statistical variation in the strength of connections disregarding their type. Remarkably, after the third test based on sex-dependent interactions, we find that our homophilic WSN network model also recovers the average weights according to edge-type, $\langle w \rangle_{FF}$, $\langle w \rangle_{FM}$, and $\langle w \rangle_{MM}$, see Fig. \ref{fig:fig2}c.

The fitted parameters in the previous example are just an instance of the various sets that satisfy the three statistical tests. Despite the fact that the nodes of the model do not have an explicit age attribute, in Fig. \ref{fig:fig3}, we show the distribution of the fitted parameters obtained by gathering different age categories to construct three empirical networks with varying group compositions. Results in Figs. \ref{fig:fig3}a-\ref{fig:fig3}b correspond to the previous association network, with A, SA, and J individuals. In Fig. \ref{fig:fig3}a, we present a comparison of the empirical measurements against the simulated results using values close to the averages (assuming a uniform prior) shown in Fig. \ref{fig:fig3}b, that is, the values of the distributions that result from the application of all significance tests (T1+T2+T3). The joint distributions of the parameters is non-trivial and in Fig. S2, we also present their distribution in parameter space. Results in Figs. \ref{fig:fig3}c-\ref{fig:fig3}d, represent the analysis considering A and SA individuals, with $(p_r,p_t,q)=(0.5,0.5,0.8)$; and in Figs. \ref{fig:fig3}e-\ref{fig:fig3}f, exclusively for A individuals, with $(p_r,p_t,q)=(0.2,0.5,0.8)$. Similar to the results in Fig. \ref{fig:fig2}, in Figs. S3-S5, we present examples of the statistical fitting for different age groups.

We interpret the higher strength of male-male connections, observed in Fig. \ref{fig:fig3}a and \ref{fig:fig3}c, as a consequence of the difference in the number of males and females in the group (as analysed in detail in the next subsection).  This pattern can be attributed to the effects of the segregation parameter $q$ and the two differentiated groups in the system (F and M), that is, male-male interactions are stronger on average because there are less possible interactions among males than among females. Note that this observation no longer holds when only adult individuals are considered, see Fig. \ref{fig:fig3}e. Although the homophilic model still fits the empirical data and recovers network features with statistical significance in that case, the empirical F-F interactions are stronger than M-M ones, despite of the fact that there are slightly more females (7) than males (6). Nonetheless, this makes sense because the majority of subadult females in the empirical system are newly immigrated and, as a result, less integrated into the group (especially with other females), exacerbating the effect of weak relationships among females.

From Fig. \ref{fig:fig3}, we can notice how the average segregation parameter, $\langle q \rangle$, is well localized for all age compositions with a small standard deviation that validates the strict inequality, $\langle q \rangle<1$, which implies that segregation effects are indeed necessary to reproduce the empirical observations related to sex-dependent interactions for all age compositions; the average triadic closure parameter, $\langle p_t \rangle$, is surprisingly uniformly spread across its all possible values for all age compositions; in contrast the average focal closure parameter, $\langle p_r \rangle$, is well localized but shifts according to the age composition. This finding is explained by looking at the difference in network sizes (as less connections are necessary when nodes of different ages are removed in the different examples) but also to the non-trivial dependency of the parameters $p_r$ and $p_t$ with $q$ (see Fig. S2, for their distribution in parameter space).

\subsection*{Numerical explorations with fitted parameters}

The previous results show that the WSN model with homophily reproduces the empirical evidence regarding stronger male-male relations ($\langle w \rangle_{MM}$) over the other types in the associations of spider monkeys. Furthermore, notice that in such empirical networks there is a asymmetrical group composition, where females represent a majority while males represent a minority group. Thus, in order to have a better understanding of the nature of our results relating to the strength of male-male connections, we considered further numerical explorations. For such simulations we considered a generic bipartisan network with nodes of type $F$ and $M$, where $F$-type nodes represent the majority group and $M$-type nodes the minority. Also, the fitted parameters, $p_r=1/2$, $p_t=1/2$, with varying homophily, $q$, network size, $N$, and percentage of the minority group, $\phi_m$. Again, no node or edge deletion is considered ($p_d=0$). 

\begin{figure}[t!]
\includegraphics[width=1.0\textwidth]{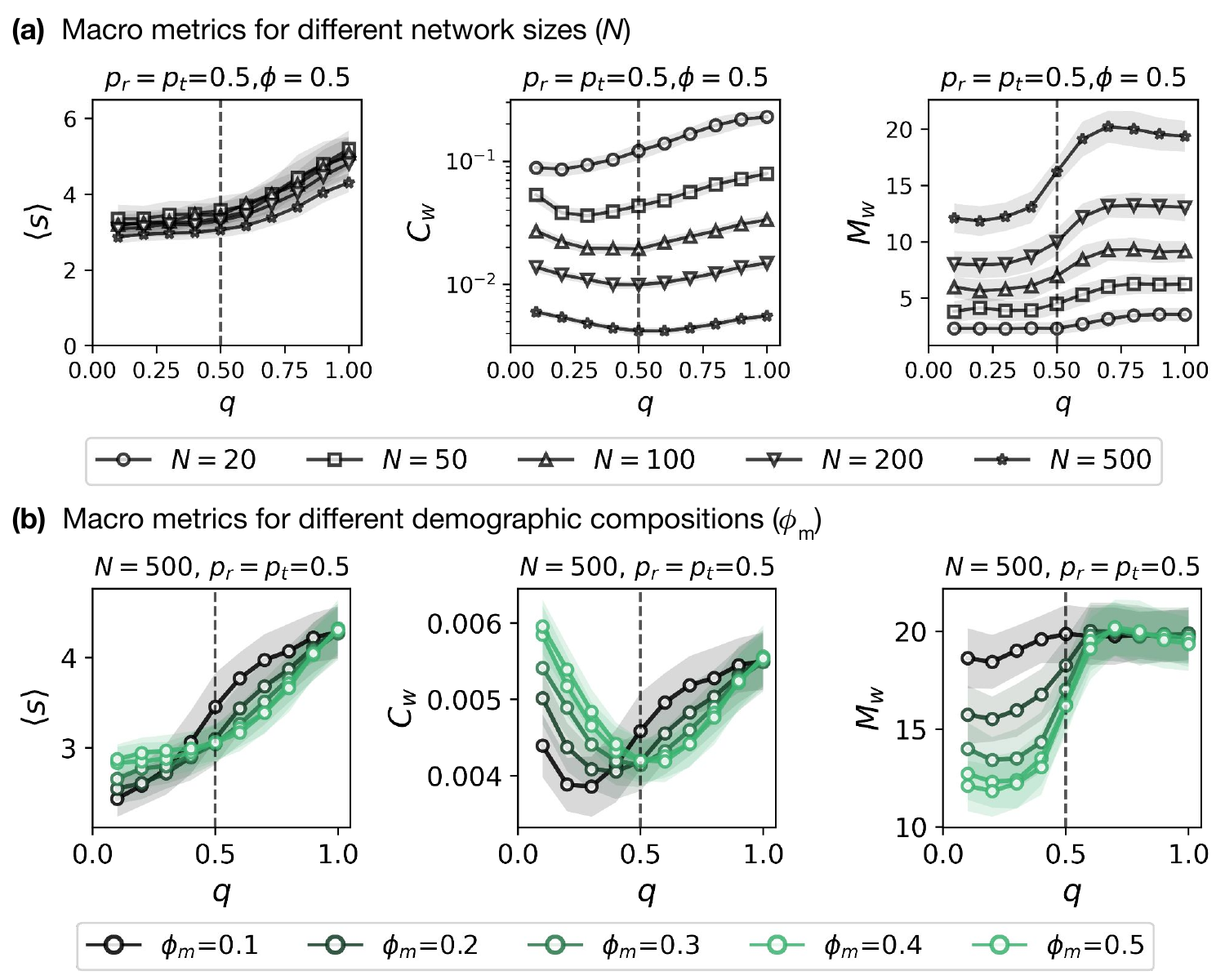}
\caption{\small \textbf{Weighted metrics for varying node-type composition and network size}. Weighted metrics, $\langle s \rangle$, $C_w$, and $M_w$, for (a) varying node-type composition, $\phi_m$, of a network of size $N=500$, and (b) varying the size of a network with $\phi_m=0.5$. In both panels, $p_r=p_t=0.5$, and $p_d=0$, the corresponding values of $\phi_m$ and $N$ are indicated by the keys, and the width of the shadowed regions is two times the standard deviation.}
\label{fig:fig4}
\end{figure}

First, to illustrate the effects of the homophily parameter $q$ on the weighted network metrics, we explored different network sizes, $N\in\{20,50,100,200,500\}$, with a fixed percentage of M-type nodes, $\phi_m=1/2$. As a direct inspection of the results shows (see Fig. \ref{fig:fig4}a), the weighted metrics display characteristic behaviors. The average strength ($\langle s \rangle$) increases monotonically with heterophily ($q \to 1$), rather independently of $N$. Meanwhile, the behavior of the number of communities ($M_w$) seems to indicate a smooth structural crossover controlled by $q$. Not only $M_w$ is clearly proportional to $N$, but is also monotonic with respect to $q$, increasing more sharply around $q=1/2$. A crossover near $q=1/2$ is also observed for the average strength, that shows two slope regimes: a nearly constant behaviour for $q<1/2$ followed by a linear increase for $q>1/2$ in Fig. \ref{fig:fig4}a.

Furthermore, the weighted clustering coefficient ($C_w$) is roughly inversely proportional to $N$, showing that smaller networks imply tighter groups, as one might expect. However, it is non-monotonic with respect to $q$, except for small networks ($N=20$). Indeed, the behavior of the weighted clustering is not trivial. In the case of large enough networks, a minimal cliquishness is attained for a certain $q\in(0,1)$. But for small networks ($N=20$), $C_w$ monotonously increases with heterophily ($q\to 1)$, making the less cliquish networks the ones that are fully sex segregated ($q=0$). This behaviour is somewhat counter-intuitive since one would expect a network divided into two small connected components ($q=0$) to be the most cliquish and reinforced among the other networks ($q>0$), where one big connected component is formed but where fewer triangles are closed. We explain this behaviour by the fact that at small $q$, less links per unit time are established and any pair of nodes with different attributes have a higher probability to fail to connect or reinforce their connection. On the other hand, at larger $q$, the same links have more opportunities to form and be reinforced, as well as triangles. This is specially true for a small network.

Second, we explored in Fig. \ref{fig:fig4}b the effects of $q$ over the weighted network metrics for different percentages of the minority group, $\phi_m=\{0.1,0.2,0.3,0.4,0.5\}$, with fixed network size, $N=500$. We chose a large network expecting it would more clearly reveal any macroscopic collective effects likely to emerge. As it can be observed in Fig. \ref{fig:fig4}b, we found that the weighted macro metrics also display non-trivial behaviors as the group composition and segregation parameters are varied: the average strength $\langle s \rangle$ increases with $q$, rather independently of $\phi_m$; the weighted clustering $C_w$ is non monotonic with $q$ and exhibits a minimum, meaning that the networks are more cliquish when fully segregated ($q=0$) or completely mixed ($q=1$) than when $q\approx 0.5$ (for $\phi_m>0.3$). Hence, an intermediate level of segregation minimises cliquishness. Noteworthy, the minimum of the weighted clustering ($C_w$) at an intermediate value of $q$ might be interpreted as a sign of structural transition, similar to the nontrivial behavior found for  the binary clustering coefficient ($C$) in a related network model with homophily (see Fig. 2b in Ref. \cite{murase2019}). In the case of Ref. \cite{murase2019}, such transition occurs at a specific number of node attributes or features ($F$), with a maximum in $C$ that increases as the number of feature values increases. In our model, a smooth structural crossover occurs at low or moderate values of the segregation parameter $q$, where $C_w$ is minimum and decreases as the proportion of the minority group decreases. Although $C$ is a measure of triangle formation while $C_w$ measures the reinforcement of such triangles in weighted networks, these extrema might be interpreted as signs of crossover transitions.

As already noticed in Fig. \ref{fig:fig4}a, the number of communities ($M_w$) also exhibits an interesting crossover transition with $q$ in Fig. \ref{fig:fig4}b. When the percentage of the minority group is small ($\phi_m=0.1$), the contribution of the latter to the community structure is negligible regardless of $q$. However, when $\phi_m$ is not so small ($\phi_m>0.3$), the number of communities increase sharply with $q$ near $q=0.5$, before reaching a plateau independent of $q$ for $q>0.5$. We interpret this increase of $M_w$ by the fact that at large $q$, each individual has effectively more possibilities of connections (outside the set of nodes sharing its attribute). We observe that during a same time period, two separated smaller groups (at $q=0$) form overall a smaller number of communities than the full group ($q=1$), indicating that $M_w$ is not additive. Together with the previous results for the weighted clustering $C_w$, the behaviour of $M_w$ provides further evidence of a structural crossover regulated by the homophily parameter $q$.

\begin{figure}[t!]
\includegraphics[width=1.0\textwidth]{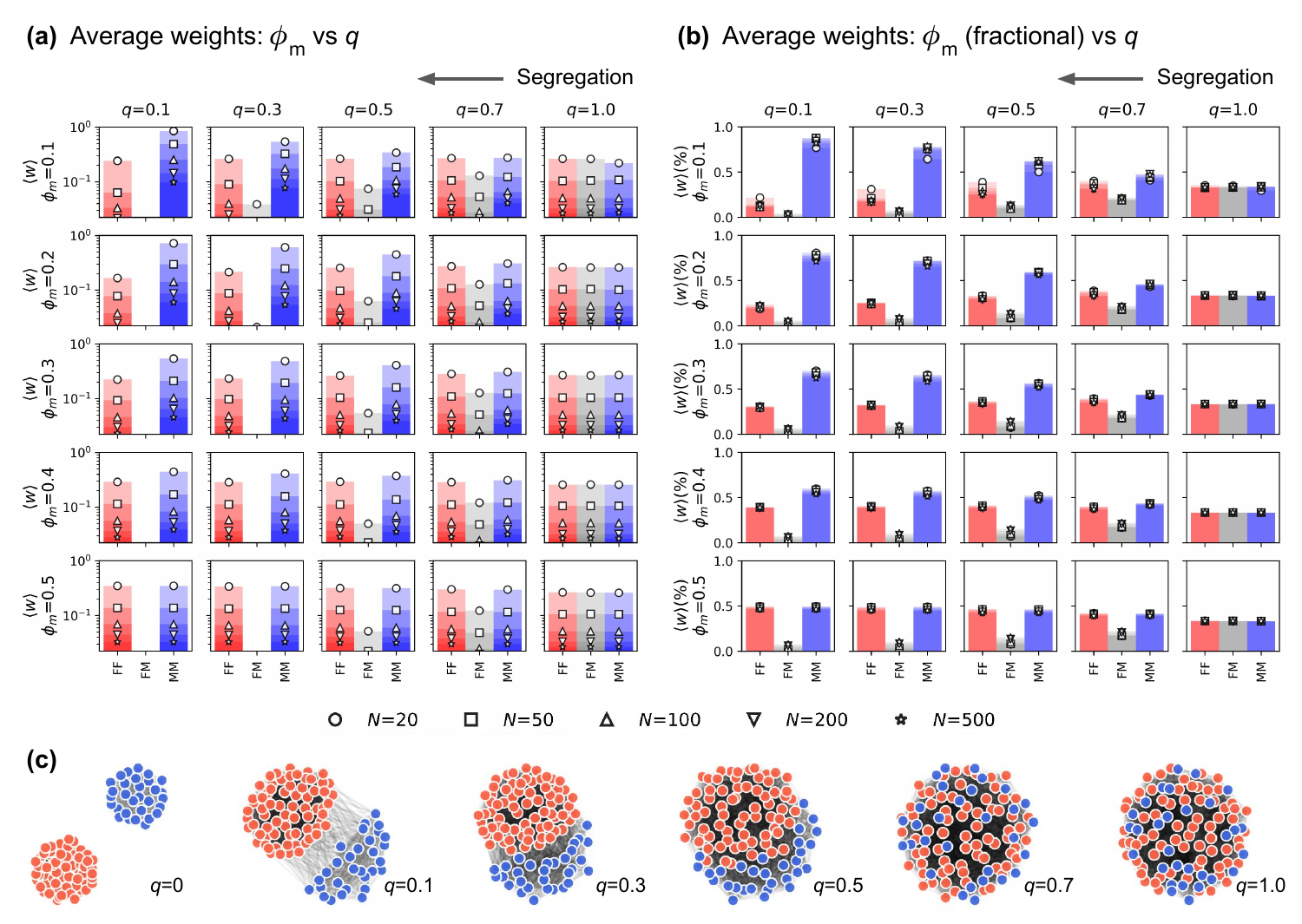}
\caption{\small \textbf{Sex-dependent weight averages in parameter space}. (a) Average weights per edge-type, $\langle w \rangle_{FF}$, $\langle w \rangle_{FM}$ and $\langle w \rangle_{MM}$, for varying demographic composition, $\phi_m$, segregation effects $q$, and network sizes, $N$, as indicated in the key at the bottom. The metrics are shown in log-scale to observe their differences. In (b) the average weights haven been normalized according to each network size. (c) An example of the networks for $N=100$, $\phi_m=0.3$, and the indicated values of $q$ (F/M-type nodes are in red/blue, respectively). In all cases, $p_r=p_t=0.5$, and $p_d=0$. }
\label{fig:fig5}
\end{figure}

Finally, in order to test the robustness of the attribute-dependent association strengths, we proceeded to systematically explore the parameter space considering the sizes $N=\{20,50,100,200,500\}$, composition with $\phi_m=\{0.1,0.2,0.3,0.4,0.5\}$, and segregation effects with $q=\{0.1,0.3,0.5,0.7,1.0\}$. In Fig. \ref{fig:fig5}a, we show the average weights per edge-type in parameter space. These average values vary according to the network size and display greater asymmetries for high segregation ($q\to 0$) than for homogeneous or fully mixed networks ($q\to 1$). It can be observed that in sex-dependent weight averages, $\langle w \rangle_{FF}$, $\langle w \rangle_{FM}$ and $\langle w \rangle_{MM}$, segregation and asymmetries in group composition tend to favor strong interactions among minorities, in this case, $\langle w \rangle_{MM}$. The interaction strength among members of the majority, $\langle w \rangle_{FF}$, vary relatively little with the segregation parameter $q$. As expected, the $\langle w \rangle_{FM}$ interactions increase with $q$ but it is quite insensitive to the composition. In contrast to $\langle w \rangle_{FF}$, the strength of minority interactions $\langle w \rangle_{MM}$ increases as segregation becomes stronger, mostly for small minorities. For a clearer visualization, in Fig. \ref{fig:fig5}b, the metrics have been divided by a normalization factor such that, for each case, $\langle w \rangle_{FF} + \langle w \rangle_{FM} + \langle w\rangle_{MM} = 1$. This normalization confirms that, indeed, the tendency to favor strong interactions among minorities (male-male interactions, $\langle w \rangle_{MM}$) is a general effect that solely depends on segregation and node-type composition ($\phi_m$) and homophily conditions ($q$), independently of the network size ($N$).

\section*{Discussion}

In this article, we presented an homophilic WSN model that is able to reproduce, with statistical significance, many of the properties of the weighted structure of an empirical association network of spider monkeys, ranging from weighted macro properties such as $\langle s \rangle$, $C_w$, and $M_w$, through the weight distribution function, $f(w)$, to the specific average weight of the sex-dependent interactions, $\langle w \rangle_{FF}$, $\langle w \rangle_{FM}$, and $\langle w \rangle_{MM}$. Notably, by considering a generic bipartisan network, in which some nodes (type $F$) represent the majority group and others (type $M$) the minority, and by performing simulations with the fitted parameters ($p_r=1/2$, $p_t=1/2$, $p_d=0$), we found that one of the main features of a spider monkey social system, namely, stronger male-male interactions over female-female or female-male ones, is a consequence of an asymmetry in the node-type composition ($\phi_m$) and homophily conditions ($q$), independently of the network size ($N$). 

Homophilic network models have been proposed to explore different aspects of social systems, such as structural transitions in networks with heterogeneous nodes \cite{murase2019}, the role of random and local connections in patterns of citation preferences \cite{bramoulle2012homophily}, the roles of race-based choice and chance in high-school friendship network formation \cite{currarini2010identifying}, or the interplay of popularity versus similarity in growing networks with preferential connections \cite{papadopoulos2012popularity}. However, the role that homophily has in the dynamical and structural phenomena related to minority groups in social networks is far from explored. For example, recent studies have shown that homophily influences the degree ranking of minorities, putting them at a disadvantage by restricting their ability to establish links with a majority group or to access novel information \cite{karimi2018homophily}, or that homophily and minority-group size explain perception biases in social networks \cite{lee2019homophily}. Notably, our general results suggest the existence of a fundamental mechanism behind the formation and reinforcement of connections among minority groups in social systems, that is, the \textsl{strength of minority ties}. In contrast to Granovetterian structures, where the strength of intra-community connections is greater than inter-community ones (which in turn define the so-called strength of weak ties), the strength of minority ties constitutes a property independent of the community structure of the network and relies solely on the type and weight of connections among nodes, group composition asymmetries, and the underlying social tie-formation mechanisms.

An example of how the strength of minority ties could be a fundamental mechanism behind social network structure is how it may be part of a simpler explanation for sex-dependent social interactions and relationships between adult spider monkeys (and other primate species with similar social behavior patterns like chimpanzees). These have traditionally been explained based on socio-ecological theory \cite{wrangham1980ecological, symington1990fission, sterck1997evolution} relating the abundance and distribution of food resources, the risk of predation, and the reproductive and social strategies evolved in each sex to improve their reproductive success. In a large-bodied primate, with relatively low risk of predation, feeding on fruit (a dispersed and variable resource), females are supposed to be subject to high degrees of feeding competition, leading to their dispersal in wide areas. Males are supposed to cooperate to defend the range of several females from neighboring groups, leading to stronger bonds between males than between females. 

Contrary to this theory, our homophilic WSN model does not assume any difference in strategies between the sexes. It retains the main mechanisms of the original WSN model \cite{kumpula2007}: the random formation of ties between two individuals under the focal closure, the reinforcement and the formation of ties with friends of friends under the triadic closure. In addition, social organization aspects of the spider monkeys' social system, such as the size and sexual composition of the group, are taken explicitly into account by considering a given number of nodes with specific attributes; the spatio-temporal cohesion is taken into account implicitly, through the tie-formation mechanisms and the dynamical evolution of the network; the interactions and relationships that comprise the social structure are considered explicitly through the tie-formation mechanisms, which in turn depended on the quality of interactions (weights) and on the attributes of the nodes (sex). Thus, the most important metrics of the empirical network can be explained by a simple bipartisan split of nodes and the previous elements.

It is noteworthy that both the GA and LA processes are necessary for reproducing the segregated structure of the empirical network. The joint distribution of the parameters is not trivial (see Fig. \ref{fig:fig3} and Fig. S2). While $p_t$ spreads uniformly over all possible values and age compositions, $p_r$ is well localized but shifts according to age composition. These findings suggest that GA dominates the creation of links (random connections) and leads faster to a fully connected network (in the absence of link deletion as in our case), rendering the role of LA as a less relevant edge creation mechanism. Somewhat surprisingly, LA favors the creation of some links (triadic connections) in the early stages of the network evolution, but its role as an edge strengthening mechanism is more relevant. Even more, the fact that $\langle q \rangle <1$ for all age compositions demonstrates the need for segregation effects to reproduce the empirical network. 

The interplay between the GA and LA mechanisms led to a raise in the weighted clustering as the size of the networks decreased, as observed in Fig. \ref{fig:fig4}a. This means that, here, only simulated networks with $N\le20$ had a large $C_w$, a feature fulfilled by nearly all social networks \cite{newman2003social}, whereas larger sizes produced more \lq\lq tree-like" structures. The social brain hypothesis \cite{dunbar1998social} actually suggests that the evolution in cognitive abilities of primates was driven by the size and number of relations in their social group \cite{boyer2018contribution}. In systems with a high degree of fission-fusion dynamics as in spider monkeys, individuals face a large uncertainty regarding their future interactions with other group members. We hypothesize that this cognitive challenge puts a limit on the group size (of the order of $20$) and that small sizes also facilitate highly transitive interactions, conferring stability to the social structure.

Furthermore, we highlight the general independence of our findings from initial conditions and configurations. Whereas our synthetic networks are created starting from just a set of isolated nodes, empirical association networks are created from connections that contain the information or memory of previous years of social interactions. And yet, the agreement between the two suggests that the application of the GA and LA mechanisms among a set of nodes with attributes generates networks that converge to structures with rather invariant properties. This provides another example of the universal aspects of the complex networks approach to the modelling of complex systems across disciplines.

Finally, further studies would benefit from a more extensive parameter search. However, the significant number of parameters of the the WSN model (i.e. the focal and triadic closure mechanisms, node/edge deletion, homophilic interactions, network size, and minority group composition) make it computationally demanding, and its output hard to analyze, which in turns hinders the fitting to empirical data. Recent advances in artificial intelligence approaches to the exploration of agent-based social network model parameters \cite{murase2021deep} represent a good alternative to deal with such difficulties by providing efficient methods for the prediction of network properties and the identification of relevant model parameters. For example, such methods could be applied to the discovery of fundamental dynamics under multiple types of interactions in a multiplex approach to social structure \cite{smith2019}. Here, not only multiple parameters drive the dynamical evolution of the social networks but also, the social structure is influenced by the endogenous and correlated effects of different layers of interacions. Investigating the multidimensional nature of social interactions in real social systems represents an attractive research direction.

\section*{Appendix: Network Metrics}

\begin{itemize}

    \item The average degree, $\langle k \rangle$, is the arithmetic mean over the nodes degree, $k_i$, defined as the number of edges connected to node $i$.
    
    \item The average strength, $\langle s \rangle$, is the arithmetic mean over the nodes strength, $s_i$, defined as the sum of the weights of all edges connected to node $i$. For the synthetic networks, the edge weights are normalized by the maximum weight of the network in order to compute the strength of the nodes and make it comparable to empirical observations.
    
    \item The average clustering coefficient, $C$, is the arithmetic mean over the nodes clustering coefficient, $C_i$, defined as, $C_i = \tau_i/\tau_{i,max} = 2\tau_i/k_i(k_i - 1)$, where $\tau_i$ is the number of pairs of neighbors involving the node $i$. The maximum number of triangles of $i$, $\tau_{i,max}$, is the number of pairs formed by the neighbors. The clustering coefficient is only defined for $k_i > 1$ (nodes with degree $k\leq 1$ are excluded of the mean). 
    
    \item The average weighted clustering, $C_w$ is the arithmetic mean of the generalized clustering coefficient, $C_{i,w}$, defined as the geometric mean of the subgraph edge weights where node $i$ participates, given by \cite{onnela2005,saramaki2007}, $C_{i,w}=[k_i(k_i-1)]^{-1}\sum_{j,k}(\hat w_{ij}\hat w_{ik}\hat w_{jk})^{1/3}$, where the edge weights are re-scaled by the maximum weight in the network, $\hat w_{ij}=w_{ij}/\text{max}(w)$.
    
    \item The number of unweighted and weighted communities or modules ($M$ and $M_w$, respectively) are computed using the Louvain community detection algorithm \cite{Blondel_2008}. 

    \item The weight distribution function, $f(w)$, is such that $f(w)\Delta w$ represents the probability that $w_{ij}\in[w,w+\Delta w]$.

    \item The average weight per edge-type, $\langle w \rangle_{FF}$, $\langle w \rangle_{FM}$, and $\langle w \rangle_{MM}$, are defined as the arithmetic mean of the edges' weights, $w_{ij}$, according to the attribute of the corresponding dyad: FF, FM, and MM. 
    
\end{itemize}

\section*{Data Availability}

The empirical network data is available from the authors of Ref. \cite{smith2019}, upon reasonable request. 

\bibliographystyle{ieeetr}
\bibliography{references}

\section*{Conflict of interests}
The authors declare that the research was conducted in the absence of any commercial or financial relationships that could be construed as a potential conflict of interest.

\section*{Authors contributions}
JRNC performed the computational analysis and wrote the main draft of the manuscript. DB defined the analytical framework and supervised the research. SESA and GRF supplemented the data and informed the spider monkeys' socio-ecological considerations. All authors participated in the discussion, writing and approval of the final manuscript.

\section*{Acknowledgments}
Grant CONACYT-CF-2019/263958 supported this work, in particular through a posdoctoral scholarship to JRNC and SEAS. The authors also acknowledge the fruitful discussions and comments from the research group that comprises the project supported by the same grant.
\end{document}